\documentclass[12pt,preprint]{aastex}

\slugcomment{\today}

\begin{document}

\title{The Substellar Mass Function: A Bayesian Approach}

\author{Peter R.\ Allen}
\affil{\small University of Pennsylvania, Dept.\ of Physics and Astronomy, 209 South 33rd Street, Philadelphia, PA 19104; pallen@hep.upenn.edu}
\author{David W.\ Koerner}
\affil{\small Northern Arizona University, Dept.\ of Physics and Astronomy, PO Box 6010, Flagstaff, AZ 86011-6010; koerner@physics.nau.edu}
\author{I.\ Neill Reid}
\affil{\small Space Telescope Science Institute, 3700 San Martin Drive, Baltimore, MD 21218; inr@stsci.edu}
\author{David E. Trilling}
\affil{\small University of Arizona, 933 North Cherry Avenue, Tucson, AZ 85721-0065; trilling@as.arizona.edu}

\begin{abstract} 
We report our efforts to constrain the form of the low-mass star and brown dwarf mass function via Bayesian inference.  Recent surveys of M, L, and T dwarfs in the local solar neighborhood are an essential component of our study.  Uncertainties in the age distribution of local field stars make reliable inference complicated.  We adopt a wide range of plausible assumptions about the rate of galactic star formation and show that their deviations from a uniform rate produce little effect on the resulting luminosity function for a given mass function. As an ancillary result, we calculate the age distribution for M, L, and T spectral types.  We demonstrate that late-L dwarfs, in particular, are systematically younger than objects with earlier or later spectral types, with a mean age of 3~Gyr.  Finally, we use a Bayesian statistical formalism to evaluate the probability of commonly used mass functions in light of recent discoveries.  We consider three functional forms of the mass function, include a two-segment power law, a single power law with a low-mass cutoff, and a log-normal distribution.  Our results show that, at a 60\% confidence level, the power-law index, $\alpha$, for the low-mass arm of a two-segment power law has a value between -0.5 and 0.5 for objects with masses between $0.04~M_{\odot}$ and $0.10~M_{\odot}$.  The best-fit index is $\alpha = 0.3\pm0.6$ at the 60\% confidence level for a single-segment mass function.  Current data require this function extend to at least $0.05~M_{\odot}$ with no restrictions placed on a lower mass cutoff.  Inferences of the parameter values for a log-normal mass function are virtually unaffected by recent estimates of the local space density of L and T dwarfs.  We find no preference among these three forms using this method.  We discuss current and future capabilities that may eventually discriminate between mass-function models and refine estimates of their associated parameter values. 
\end{abstract}

\section{Introduction}

The Initial Mass Function (IMF) is one of the fundamental distributions in modern astronomy.  The IMF, ${\Psi}(m)$, describes the number of stars born per unit mass, per unit volume.  The concept was introduced by \citet{salp}, who characterized the distribution as a power law, $\Psi(m) \propto M^{-\alpha}$, with $\alpha = 2.35$.  Salpeter's analysis extended nominally to $\sim0.3~M_{\odot}$, but included relatively few low mass stars.  Two decades later, \citet{ms79} used improved observations of lower luminosity stars to show that the mass function flattens at lower masses.  Subsequent analyses have generally characterized the IMF as either a combination of power laws (with $\alpha \sim 1$ below $\sim1~M_{\odot}$ and close to Salpeter's value at higher masses) or as a log normal distribution, $\Psi(m) \propto \exp{(\frac{\log(m) - \log(m_0)}{\sqrt{2}\sigma})^{2}}$, as in Miller \& Scalo.  As yet, no direct connection has been uncovered between either functional form and the underlying physical mechanisms of star formation.

The measurement of the substellar mass function has great implications for the theory of star formation.  Many different theories for the formation of brown dwarfs have been proposed to reproduce the observed frequency of brown dwarfs as field objects, companions, and members of young clusters.  \citet{kru03} provide a good review of the different formation arguments.  At present, there are two main paradigms for brown dwarf formation: fragmentation of molecular clouds, the same process that forms stars \citep{bri02,wb03}; and ejection of pre-stellar embryos \citep{rc01,dcb03,sd03}.  In general, the first paradigm predicts more brown dwarfs, as a continuation of the stellar IMF, while the second predicts fewer brown dwarfs, resulting in a sharp drop in the mass function near the hydrogen burning minimum mass.

Recent studies of the stellar IMF, summarized in \citet{scalo96}, \citet{scalo98}, \& \citet{kru03}, have shown a similar IMF in many different environments, including the local Galactic field and many different star formation regions.  However, only in recent years has it become possible to probe the low-mass star and substellar regime to estimate the underlying mass function.  The majority of these analyses have centered on young ($<50$ Myr) star clusters. In principle, the higher luminosities of young brown dwarfs in these clusters allow the mass function to be derived for masses as low as 5-15 Jupiter masses, although the theoretical models used to calibrate the observations become increasingly uncertain at these extremely young ages \citep{bar02}.  Nevertheless, the available measurements can be characterized using either the power law or log normal formalisms.  Table 1 provides a representative set of results from these studies, characterizing the substellar mass function using the power law index $\alpha$.  All of the cluster values have $\alpha < 1$, indicating a flatter mass function (lower space densities) than that associated with field M dwarfs (e.g. \citet{rg97} found $\alpha \sim 1$). This result is also highlighted by \citet{kru02}, whose value of $\alpha$ for the average solar neighborhood is based predominantly on results from young star clusters. 

Few direct studies of the substellar IMF in the field have been made, and Table 1 lists their results.  The value of these derived power law exponents are generally greater than or equal to those obtained from cluster studies. However, it is important to bear in mind that the mass range sampled in the field is much more restricted than in the clusters.  Moreover, a prime scientific driver of some M and L dwarf studies is to test whether or not the brown dwarf frequency in the field is sufficient to contribute significantly to local dark matter.  It is now generally accepted that the low-mass star and brown dwarf mass function is incompatible with an extension of the Salpeter slope, and that brown dwarfs do not consititute an appreciable fraction of the local dark matter density (emphasized particularly in \citet{inr022}). 

\citet{chab03} has recently reanalyzed the available space density data for stars and brown dwarfs in the field.  He favors characterizing the IMF as a log normal with $m_0 = 0.08~M_\odot$, rather than the \citet{ms79} value of $m_0 = 0.15~M_\odot$. Conceptually, the log normal distribution is equivalent to a set of power laws where $\alpha$ decreases with decreasing mass. Thus, a log normal distribution can match the number of brown dwarfs predicted by a moderately steep power law ($\alpha \sim 1$) at $m > 0.05~M_{\odot}$, but it predicts substantially fewer brown dwarfs at lower masses than a single power law.

We explore the substellar mass function through extensive modeling and statistical analysis of the available data for the nearby field population of late-M, L, and T dwarfs.  \citet{burg04} has recently undertaken a similar study using Monte Carlo techniques and will be compared to our results; \citet{mll} have probed similar issues through observations of young clusters and associations.  The data for the present study are compiled from a volume-limited sample of late-M and L dwarfs \citep{kc03,kc05}, combined with initial estimates of the local space density of late-T dwarfs \citep{burgp}.  We refer to this compilation as the KCAB dataset.  We will examine three different models of the underlying substellar mass function.  We first consider a two segment power law:
\begin{equation}
{\Psi}(m) \propto \left\{ \begin{array}{c} m^{-1.05}, m \geq m_{12} \\ m^{-{\alpha}_2}, m < m_{12} \end{array} \right\}
\end{equation}
where $m$ is the mass in solar units, and $\alpha_2$ and $m_{12}$ are the power law exponent and the segment joining mass, respectively, as defined in \citet{a03} (A03).  The value of $m_{12}$ is limited by the mass range of the \citet{bur} models, which we use to construct our model luminosity functions, to values between 0.001~$M_{\odot}$ and 0.15~$M_{\odot}$.  The power law exponent for masses greater than $m_{12}$ is set equal to the value determined by \citet{rg97}.  We next consider a single power law with a low-mass cutoff:
\begin{equation}
{\Psi}(m) \propto \left\{ \begin{array}{c} m^{-\alpha_2}, m \geq m_{cut} \\ 0, m < m_{cut} \end{array} \right\}
\end{equation}
where $m_{cut}$ is the cutoff mass in solar masses.  This form was suggested by \citet{kru03}.  They determined that an abrupt end to the low-mass star mass function was consistent with the data.  Finally, we examine a log normal distribution:
\begin{equation}
\Psi(\log(m)) \propto \exp{\left\{ \frac{(\log(m) - \log(m_0))^{2}}{2{\sigma}^{2}}\right\} }
\end{equation}
where $\log(m_0)$ gives the center of the distribution, and $\sigma$ controls the width.  A recent log normal fit to objects between a few $M_{\odot}$ and the stellar/substellar limit yields $\log(m_0) = -1.1\pm0.1$ and $\sigma = 0.69\pm0.05$ \citep{chab03}.  In this paper we compare the predictions of these models to the data using a Bayesian statistical approach.  The Bayesian method provides a rigorous and elegant statistical comparison between disparate datasets and models \citep{press}.

The present paper is organized as follows.  Section 2 reviews our derivations of theoretical luminosity functions, as previously discussed in A03, and examines the morphology of the new synthetic field luminosity functions.  Section 3 describes the KCAB dataset and the Bayesian statistical method, including tests and final analysis.  Section 4 considers the ways in which future observations can better constrain the field IMF. Finally, Section 5 summarizes our conclusions.

\section{Model Field Luminosity Functions}

The study of the substellar IMF is more complicated than the stellar IMF because, unlike stars, brown dwarf temperatures and luminosities evolve rapidly as a function of time.  Moreover, brown dwarfs of different masses follow almost identical tracks in the HR diagram and evolve through the same sequence of observable features regardless of mass.  This leads to a degeneracy between mass and age; for example, a mid-type L dwarf could be a several Gyr-old 0.07~M$_\odot$ object, or a young ($<50$ Myr) $\sim0.02~M_\odot$ object. There are two methods of handling this degeneracy: surveys of young stellar clusters, where age is a known parameter, and statistical analysis of the field distribution.  A03 studied the young clusters whereas the paper covers the statistical inference of the field IMF.

The brown dwarf age/mass degeneracy means that there is not a unique transformation between the observed luminosity function and the underlying mass function for a mixed age population.  Instead, we invert the transformation.  Starting with mass and age distributions, we derive theoretical luminosity functions via the \citet{bur} evolutionary models of low-mass dwarfs, the same models used in A03 (see that paper or \citet{burg04} for a comparison with other evolutionary models).  By varying the underlying physical distributions, we obtain the combination of mass and age distribution that best reconstructs the observed luminosity function.

\subsection{The Age Distribution of the Field}

The age distribution of stars in the local Solar Neighborhood is difficult to determine because most age indicators work best at young ages ($<$ 1-2 Gyrs).  Estimates have been derived using a variety of techniques, including analysis of the distribution of chromospheric activity in nearby G- and M-dwarfs \citep{sod,pmsu3}; modeling the color-magnitude diagram in the Solar Neighborhood \citep{hern}; and using observations of distant galaxies to infer star formations rates as a function of redshift \citep{pas}.  Among these studies, only \citet{sod} explicitly cite uncertainties in the derived star formation rates; we take those values ($\sim20\%$) as characteriestic of this type of analysis.

Figure \ref{fig:fad} compares four representative empirically-derived age distributions for field stars against three idealized distributions.  Each distribution indicates the probability of a field star having a particular age.  Most empirical measurements, despite large apparent excursions, are broadly consistent with a constant star formation rate.  We use three idealized cases to test the effects that the underlying age distribution has on the output luminosity functions and $T_{eff}$ distributions.  The idealized age distributions are as follows: decreasing star formation (fewer stars formed today than early in the galaxy's history), with a slope of 0.02 Gyr$^{-1}$; increasing star formation (more formation today than 10 Gyr ago), with a slope -0.02 Gyr$^{-1}$; and uniform star formation. All age distributions span 0 to 10 Gyr, with 0 as the present and 10 as 10 billion years ago.  As illustrated in Figure \ref{fig:fad}, these three idealized models cover the extreme range of trends suggested by the empirical estimates.

\subsection{Bolometric Luminosity Functions and $T_{eff}$ Distributions}

The various IMF forms and assumed age distributions, described in the previous sections, are combined to create continuous distributions of objects as a function of mass and age.  Each model mass--age distribution is normalized such that the space density of 0.10~M$_{\odot}$ objects is equal to 0.35 stars pc$^{-3} M_\odot^{-1}$ \citep{rg97}.  Finally, the \citet{bur} low-mass star and brown dwarf evolutionary models are used to assign values of bolometric luminosity and effective temperature at each point in the mass--age distribution.  Luminosity functions and $T_{eff}$ distributions are derived from this distribution by summing the number of objects in an interval of luminosity or $T_{eff}$.

The substellar nature of brown dwarfs leads to characteristic structure in the luminosity and $T_{eff}$ distributions.  As a baseline reference, we use a two-segment power law mass function with $\alpha_2 = 1.0$ and $m_{12} = 0.09~M_{\odot}$ (essentially a continuation of the low-mass star IMF) throughout the paper unless otherwise noted.  Figure \ref{fig:asc} shows the luminosity functions and $T_{eff}$ distributions derived for the baseline model coupled with the three idealized age distributions described in Section 2.1.  In each case, the bolometric luminosity function exhibits three prominent features: a rise to bright magnitudes (A), a deep trough (B), and a large clump at faint magnitudes (C).  Feature A is predominantly main sequence stars, together with a small number of young brown dwarfs; the sharp cutoff at high luminosities stems from the upper mass limit of the Burrows models ($0.15~M_{\odot}$). Brown dwarfs account for almost all sources with M$_{bol} > 12$, and their characteristic evolution accounts for features B and C.  The rate of cooling and consequent fading in luminosity increases with decreasing mass and age while, for all masses, it decreases with decreasing temperature. Brown dwarfs decrease rapidly in luminosity and $T_{eff}$, which leads to relatively small numbers of luminous brown dwarfs, creating trough B. As the rate of cooling decreases, for older brown dwarfs, objects accumulate at lower luminosities, leading to feature C.  Note that the cutoff after feature C is due to the low mass edge of the Burrows models.  Overall, the luminosity functions in Figure \ref{fig:asc} have similar morphologies to those computed for young clusters in A03 (see Figure 3 of that paper). The main difference is the absence of the transient Peak D, a spike in the luminosity function produced by brown dwarfs that burned deuterium. Peak D does not appear in the field luminosity function because the solar neighborhood includes few very young brown dwarfs.

The general shape of the $T_{eff}$ distribution is similar to that of the bolometric luminosity function (Figure \ref{fig:asc}b).  The effective temperature regimes marked for M, L and T spectral types are based on those given by \citet{golim}.  Trough B is centered in the L dwarf regime, $\sim$2300K--1450K.  As noted above, the substellar objects in that effective temperature range evolve particularly rapidly (see Figure 8 of \citet{bur}). Moreover, the models indicate that only stars less massive than $m < 0.082~M_\odot$ enter the L dwarf regime, while the coolest hydrogen-burning objects reach temperatures of $\sim1800$K (spectral type L3/L4).  Thus, the L dwarf population is composed of brown dwarfs that spend very little of their lifetimes as L dwarfs and main sequence stars drawn from a very limited mass range. 

There are only limited differences in the predicted luminosity functions despite the widely-varied age distributions used in their construction.  Those differences are restricted to two regions: $15 < M_{bol} < 18.5$, or late-L at T dwarfs; and $M_{bol} > 25$, well beyond known T dwarfs.  The increasing star-formation model predicts more young objects, with correspondingly higher numbers of L and early-T dwarfs, while the decreasing star-formation model predicts more older objects and larger numbers of late-T and cooler dwarfs.  Given the results shown in Figure \ref{fig:asc}, and current observational uncertainties, we adopt a uniform age distribution (constant star formation rate) as the reference distribution throughout the rest of the present study.

Figure \ref{fig:sfmf} shows the predicted $M_{bol}$ distributions for three power law IMF models, with $\alpha_2$ set to 0.0, 0.5, and 1.0.  Note that the variations through the L dwarf regime are smaller than the fluctuations produced by varying the star formation history (Figure \ref{fig:asc}), while there are more substantial, systematic changes in the predicted T dwarf number densities.  We therefore conclude, {\sl a priori}, that L dwarf luminosity function data are unlikely to provide strong constraints on the underlying mass function. T dwarf data are essential to obtain a reliable estimate of the substellar mass function.

The synthetic luminosity and $T_{eff}$ distributions produced by \citet{burg04} were generated using a different statistical approach (Monte Carlo), but with similar evolutionary models \citep{bur97} and assumptions on the IMF and age distribution.  Consequently, there are few differences between Burgasser's models and ours.  Burgasser finds similar responses in the luminosity function to variations in the age distribution as found in this paper.  In Burgasser's Figure 10, several model luminosity functions and $T_{eff}$ distributions are displayed.  His models show the same decrease in space density for mid-L through early-T dwarfs for an age distribution weighted toward old objects as seen in our Figure 2.  Changes in the power law index $\alpha$ also produce similar results between these two works.  In Figure 4 of Burgasser, we see model bolometric magnitude luminosity functions for several values of $\alpha$, which is similar to our Figure 3.  The location of the bottom of trough B is consistent between the figures at M$_{bol} \sim 14.5$.  Additionally, the size of the low-mass end of the luminosity function (C) grows with increasing $\alpha$ in a similar way between the two works.  The good agreement between our analysis and Burgasser's independent analysis suggests that these results are representative of current theoretical models.

\subsection{Broad-band Luminosity Functions and Bolometric Corrections}

Empirical surveys derive a luminosity function either in a particular bandpass ($I_{c}, J,  $or $K$), or as a function of spectral type (an observational surrogate for T$_{eff}$) not the bolometric luminosity provided by the evolutionary models.  Consequently, the predicted bolometric luminosity functions must be transformed to broad-band luminosity functions for comparison with observations.  We have adopted the bolometric corrections of \citet{golim} to transform the bolometric luminosity function to $K$-band and $M$-band.  Figure \ref{fig:bck} compares the \citet{golim} $K$-band bolometric corrections against the relation adopted in A03, based on data from \citet{dahn}.  We have used $IJH$ colors for late-M, L and T dwarfs from several sources {\citep{leg02,dahn,legm02,knapp04} to derive bolometric corrections in those bands.  The same methods are employed here as in A03 to carry out the conversion using the newer bolometric corrections.  

Figure \ref{fig:bblfs} compares the field bolometric luminosity function for a uniform age distribution to the corresponding broad-band luminosity functions.  The overall morphology is generally preserved, with the main differences lying in the magnitude range spanned by the luminosity function and the width and depth of trough B.  The one distinct difference from A03 to this work is that a feature in the $K$-band luminosity function at M$_K=14$, which we attributed to the onset of methane absorption, has disappeared.  This feature was introduced by a kink in the A03 bolometric corrections close to the L/T transition.  The new \citet{golim} results provide a smoother relation through the transition, removing the extra dip (see Figure \ref{fig:bck}).  Given the absence of empirical data at M$_J > 16$, we are forced to extrapolate the bolometric corrections to temperatures lower than $\sim 700~K$ (spectral types later than T8).  Thus, these features have larger uncertainties than brighter sections in the broad-band luminosity functions at the corresponding absolute magnitudes (marked by an arrow in Figure \ref{fig:bblfs}).

\subsection{Age and Mass Distributions as a Function of Spectral Type}

In general, the age distribution for a particular substellar spectral type will not be flat even if a uniform star-formation rate is assumed.  The theoretical models discussed here can be used to predict mass or age distributions of objects as a function of spectral type.  Figure \ref{fig:spta} shows illustrative results from our nominal model.  We show the predicted probability density distributions (i.e., the likelihood, per unit mass or age, an object has a certain mass or age) for spectral types M6, L0, L5, late-L, early-T, and late-T.  The top panel of Figure \ref{fig:spta} plots the age distributions and the bottom panel the mass distributions, and Table 2 lists the average age and mass of each spectral type.  Given the uncertainties inherent in the models at very young ages, we have discarded all model results with ages less than 20 Myrs (0.2\% of the sample for a constant star formation history).

We examine the features of these mass and age distributions to better understand the physical properties of the underlying substellar population.  The M6 age distribution is essentially flat. This reflects the overwhelming predominance of hydrogen-burning stars.  In our baseline model, M6 dwarfs form at a uniform rate over the 10~Gyr spanned by the simulations and settle rapidly onto the main sequence with little subsequent evolution.  In contrast, the relative proportion of young ($< 2$ Gyr) dwarfs increases, and the average age decreases, as one progresses down the L and T dwarf spectral sequence (see Table 2).  This behavior stems partly from the decreasing contribution of hydrogen-burning stars and partly from rapid cooling of brown dwarfs through these temperature regimes.  Late-type T dwarfs, however, exhibit a much flatter age distribution, albeit still decreasing with increasing age.  The constant birthrate of new brown dwarfs, coupled with the slower cooling rate at these temperatures ($\sim$1250K--700K), as compared to L dwarfs, yields an approximately constant density of late-T dwarfs as a function of age.  This is reflected in the average age of $\sim 5$ Gyr, comparable with that of stellar-mass M6 dwarfs. 

The mass distribution as a function of spectral type changes significantly as one crosses the stellar/substellar boundary.  As can be seen in the bottom panel of Figure \ref{fig:spta}, the mostly stellar M6 mass distribution has a well-defined mass range with a shallow tail.  The L0 and L5 mass distributions are strongly peaked at masses near the hydrogen-burning limit, which reflects the long main sequence lifetimes of stellar L dwarfs.  The distribution broadens for late-L and early-T dwarfs, both of which include only substellar-mass objects.  Nevertheless, higher-mass brown dwarfs, which spend long periods of time as L and early-T dwarfs, are the majority constituent in both cases.  The average mass is lower for early-T dwarfs, since the highest-mass brown dwarfs ($m > 0.055~M_{\odot}$) in the Galactic disk have not had sufficient time to cool to temperatures below $\sim1300$K.  Much longer cooling times in the late-T temperature range lead to a very broad mass distribution, although one should note that most of the lowest-mass brown dwarfs have dropped below T$_{eff} \sim 700$K, the lower temperature limit for this bin.

\section{Bayesian Inference of the Field Mass Function}

\subsection{The Bayesian Approach}

In developing and presenting a Bayesian approach to the study of the mass function we hope to encourage the application of more rigorous statistical methods to this field.  As noted by \citet{press}, the state of most statistics approaches to astronomical data analysis is lamentably simple.  Using standard `freshman lab' statistics one can overemphasize a result because of incorrect uncertainties or be baffled by apparently incompatible datasets.  A Bayesian approach is able to cope with such problems and to quantify the relative degree of belief of one model over another.  We use this method here to evaluate inferences about the substellar mass function based on recent observations.

The core of this method is Bayes' rule:
\begin{equation}
P(\theta|D) \propto P(D|\theta){\times}P(\theta)
\end{equation}
where $\theta$ is the model, $D$ is the data, $P(\theta|D)$, the {\it posterior distribution}, is the probability of the model given the data, $P(D|\theta)$, the {\it likelihood function}, is the likelihood of the data given the model, and $P(\theta)$, the {\it prior distribution}, is the initial probability of the model \citep{siv}.  The output posterior distribution provides a wealth of information on the model parameters, from the best fit parameter set to correlations between parameters.  Using Bayes' rule we calculate the probability that the data would have been measured given a hypothesized model (the likelihood function).  However, we wish to know the probability that a hypothesized model is true given the measured data (the posterior distribution).  The power of Bayes' rule lies in the simple relation of these two quantities.

The specification of a prior distribution remains the most controversial aspect of Bayesian analyses and must be considered carefully.  The prior folds previous observational and theoretical evidence into the analysis in more than one manner.  One technique uses the posterior distribution from a previous analysis.  This enables the same analysis to be performed in light of new and improved data.  In this way, one can iterate over multiple datasets thereby incorporating them into the analysis of a single set of models to provide one unified result.  This is ideal for the study of the field substellar mass function because there is no single dataset for all low-mass stars and brown dwarfs, and new data is continually made available.  Priors can also be constructed if no previous knowledge of the problem exists.  In this case the prior distribution should not impart a bias on any parameter value under consideration.  These types of priors fall under the broad heading of conjugate distributions.  The prior distributions used here will be constructed from previous estimates of the substellar mass function.

\subsection{Check the Method}

To check that our Bayesian method is correctly implemented, we attempt to reproduce the results of \citet{inr022} (hereafter RGH) for the nearby stellar mass function.  RGH compiled observations of M dwarfs from catalogs, supplemented the available distance estimates with Hipparcos parallaxes, included Hipparcos data on G and K stars, and selected all objects within specific distance and absolute magnitude limits.  Those data, coupled with an empirically derived mass-luminosity relation, enabled them to fit the stellar mass function from 1~$M_{\odot}$~to $\sim0.1~M_{\odot}$.  We take their mass estimates and perform our own analysis using a Bayesian method.

To begin, a model must be constructed to compare with the data.  We use a power law mass function model with only one free parameter, $\alpha$, the power law index.  We set the resolution to $\Delta{\alpha} = 0.01$ and allow $\alpha$ to vary from 0.0 to 2.0, consistent with the range of values given in Table 1.  Each computation is normalized to match the density given by the sum of the three highest RGH mass function bins.  The likelihood and prior distributions also need to be specified.  Since the RGH data set is well defined and contains large numbers a standard Gaussian functional form is used for the likelihood function \citep{siv}:
\begin{equation}
P(D_i|\theta_i) \propto \exp{\left\{ -\frac{(\theta_{i}-D_i)^2}{2{\sigma}_{i}^2}\right\} }
\end{equation}
where $D_i$ is the $i$th bin in the measured mass function, $\theta_i$ is the $i$th bin in the model mass function, and $\sigma_i$ is the uncertainty in the measurement of the $i$th data point.  The prior distribution used is based on the average of the estimates of $\alpha$ listed in Table 1 and their uncertainties, and is given the shape of a Gaussian:
\begin{equation}
P(\alpha) = \frac{1}{\sqrt{2{\pi}}{\sigma}_{\alpha}}\exp{\left\{ -\frac{({\alpha}-{\alpha}_0)^{2}}{2{\sigma}_{\alpha}^{2}}\right\} }
\end{equation}
where $\alpha_0$ = 0.8 and $\sigma_{\alpha}$ = 0.9.  The calculation to determine the posterior distribution can now be carried out, and has the following functional form:
\begin{equation}
P(\alpha|D(RGH)) \propto \prod_{i} \exp{\left\{ -\frac{(\theta_{i}(\alpha)-D_i)^2}{2{\sigma}_{i}^2}\right\} }{\times}P(\alpha)
\end{equation}
where the product is over every bin in the mass function.  We assume that each data point is independent.  This allows the individual $i$th distributions to be multiplied to obtain the final posterior distribution for $\alpha$ \citep{siv}.

We find a best-fit power law mass function with $\alpha = 1.09 \pm 0.017$, consistent with the value of 1.15 $\pm$ 0.2 derived directly in RGH.  The best-fit model mass function is displayed with the data in Figure \ref{fig:pf}a, and the resultant Bayesian posterior distribution, with the input prior distribution, for $\alpha$ is shown in Figure \ref{fig:pf}b.  We see that the posterior distribution is Gaussian, is centered at a significantly different value than the prior distribution (1.15 instead of 0.8), and is more tightly constrained than the prior.  Since we have reproduced the earlier results and the resultant posterior distributions are well-behaved, we believe that our Bayesian method is correctly constructed.  We will proceed to use it to understand the substellar mass function.

\subsection{Model Fits to the Substellar Mass Function}

We now extend the above demonstration of the utility of a Bayesian approach to include the substellar mass function of the local field.  As described in the Introduction, we use number counts of M7 to L8 dwarfs taken from \citet{kc03,kc05}, and space densities of T5-T8 dwarfs taken from \citet{burgp} to study the field substellar mass function.  We combine those data into a joint $J$-band luminosity function in order to compare them to our models.  \citet{kc05} provides a $J$-band luminosity function directly (although known to be incomplete for M7 dwarfs (M$_J < 11$)).  However, the \citet{burgp} T dwarf data provide space density as a function of spectral type.  We have used the T dwarf $M_J$-Spectral Type relation from \citet{vrba} to transform the spectral type distribution to a $J$-band luminosity function:
\begin{equation}
M_J = 15.04 - 0.533{\times}SpT + 0.091{\times}SpT^2
\end{equation}
where $M_J$ is the absolute $J$-band magnitude and $SpT$ is a spectral type index (T0-T8 = 0-8). Combining those results with the Cruz et al.\ data gives the empirical KCAB $J$-band luminosity function listed in Table 3.  

The M$_J$-Spectral Type relation is double-valued for spectral types between $\sim$L5 and T5, reversing its direction at around L/T transition, with early-type T dwarfs and late-type L dwarfs having similar M$_J$ \citep{vrba}. Survey data for this absolute magnitude regime ($14.0 \le M_J \le 15.5$) are incomplete. This reflects both known incompleteness in the \citet{kc05} survey for L5-L8 dwarfs and the absence of published density estimates for T0-T4 dwarfs. Thus, the space densities between $M_J = 14$ and $M_J = 15.5$ listed in Table 3 are lower limits, and these data will not be used to constrain the field substellar mass function.

As outlined in the introduction, three different mass function models will be used to fit to the data: a two-segment power law (as used in \S2 and A03), a log normal mass function, and a single power law with a low-mass cutoff.  The segmented power law \citep{kru03,inr99} and the log normal distribution \citep{chab03} mass functions have long standing traditions.  The third formulation is chosen because the field data may be consistent with a mass function that continues through the stellar/substellar boundary and then abruptly cuts off \citep{kru03}.  Therefore, we consider this formulation and compare the resultant luminosity function to those generated by the two more standard forms of the mass function.

\subsubsection{Two Segment Power Law Mass Functions}

As with the test case, we must first set range of the model parameters to be probed and construct functional forms for the distributions on the right hand side of Equation 4.  For the two segment power law, as described in \S2, the modeled region is -1.5 $\le$ $\alpha_2$ $\le$ 1.5 both to sample the range given in Table 1 and to allow for a sharp drop in the mass function.  The value of $m_{12}$ is limited to be 0.01$M_{\odot} \le m_{12} \le 0.1M_{\odot}$, lying in the range of masses potentially probed by the KCAB dataset and within the \citet{bur} models.

The prior distributions for $\alpha_2$ and $m_{12}$ are straightforward.  The $\alpha_2$ prior discussed in \S3.2 is used again, since it is based on empirical estimates (Table 1).  However, there is no previous knowledge about $m_{12}$, so care must be taken not to impart a bias on the posterior distribution.  We use a maximum entropy argument to determine the prior distribution on $m_{12}$.  To do so we need to define the constraints to which the distribution on $m_{12}$ is subject.  The only constraint on the probability distribution of $m_{12}$ is an invariance to changes in scale; i.e., the units of mass can be changed from solar masses to Jupiter masses with no effect on the outcome.  This means that $m_{12}$ is a scale parameter, and the most `ignorant' prior distribution is given by $P(m_{12}) \propto 1/m_{12}$ \citep{siv}.  Therefore, the 2D prior distribution is given by the following:
\begin{equation}
P(\alpha_2,m_{12}) \propto \frac{1}{\sqrt{2{\pi}}{\sigma}_{\alpha}}\exp{\left\{ -\frac{({\alpha}-{\alpha}_0)^{2}}{2{\sigma}_{\alpha}^{2}}\right\}}{\times}\frac{1}{m_{12}}
\end{equation}

The KCAB dataset is limited by small numbers as there are no more than 15-20 objects in the largest magnitude bin, with most having less than 10.  Consequently, a Poisson form will be used for the likelihood function, unlike in $\S$3.2.  A Poisson distribution is best suited for small numbers of objects \citep{siv}.  The Poisson form is given by: $P \propto Rate^{N_{det}}{\times}\exp{(-N_{obs}{\times}Rate)}$, where $Rate$ is the predicted model space density; $N_{det}$, the detected number of objects, is the product of the observed density and the volume searched; and $N_{obs}$, the number of observations, is the volume observed.  Hence, the likelihood function is given by the following form:
\begin{equation}
P(D_{i}|\theta_{i}) \propto \theta_{i}(\alpha_2,m_{12})^{D_{i}V_{i}}\exp{(-V_{i}\theta_{i}(\alpha_2,m_{12}))}
\end{equation}
where the subscript, $i$, represents each absolute $J$ magnitude bin in the luminosity function; $D_i$ and $V_i$ are the measured space density and the volume explored in each magnitude bin of the KCAB dataset, respectively; and $\theta_i$ is the model space density for the $i$th $J$ magnitude bin.  The final unnormalized posterior distribution is given as the natural log of the product over each magnitude bin of Equation 10:
\begin{equation}
{\ln}(P(\alpha_2,m_{12}|D)) \propto {\ln}(P(\alpha_2,m_{12})) + \sum_{i} D_{i}{\times}V_{i}{\times}{\ln}(\theta_{i}(\alpha_2,m_{12})) - V_{i}{\times}\theta_{i}(\alpha_2,m_{12})
\end{equation}
The prior distribution is outside the summation because it is invariant of the specific data point under consideration.

To calculate the posterior distribution, we generate a series of mass-age distributions and transform them to the observational plane via the brown dwarf models, as described in $\S2$.  The resolution of the mass-function model parameters are $\Delta{\alpha_2} = 0.05$ and $\Delta{m_{12}} = 0.001~M_{\odot}$.  We normalize each iteration to match the space density for objects with $M_J$ = 11--12.5 in the KCAB luminosity function.  These data sample the most luminous ultracool dwarfs and are likely to provide the most reliable space density estimates.  

Figure \ref{fig:ka4} displays the resultant 2D posterior distribution for $\alpha_2$ and $m_{12}$.  The most probable solution is $\sim \alpha_2 = 0.0$ and $M_{12} \sim 0.08~M_{\odot}$ with large uncertainties.  We also test the effect that varying the $\alpha_2$ prior distribution has on the output posterior distribution.  Figure \ref{fig:twoplp} displays four 1D posterior distributions derived from four prior distributions for $\alpha_2$; the nominal case, a shifted case, a wider case, and a narrower case.  The overall shape and peak location of the posterior distribution remains largely the same despite the variations of the prior distribution.  The narrow prior distribution produces the biggest differences.  However, all the posterior distributions cover similar ranges in $\alpha_2$ with similar amplitudes.  Although the posterior distributions differ substantially from their priors, they are still affected by them and only weakly constrain model parameter values.  Consequently, the different output posterior distributions all fit the data equally well.

\subsubsection{Cutoff Power Law Mass Functions}

The choice of a two-segment power law is not clearly required, so we also fit the data with other forms of the mass function in an effort to determine which, if any, provides the best fit to the KCAB dataset.  \citet{kru03} suggest that the field mass function may be consistent with a cutoff (a steep drop in number density) at or near the stellar/substellar boundary.  We test this hypothesis using our Bayesian formulation, which uses identical likelihood functions and prior distributions as those given in \S3.3.1, but with new mass function models given by Equation 2.  Figure \ref{fig:ka5} displays the resultant 2D posterior distribution.  The maximum is at the lower edge of the $m_{cut}$ range, $0.01~M_{\odot}$, with $\alpha_2 \sim 0.25$.  Our results agree better with the \citet{kru03} mass function that includes no lower mass cutoff, rather than a cutoff near the hydrogen burning limit; $0.05~M_{\odot}$ is the highest cutoff mass that is consistent with our analysis.  This upper limit is approximately the lowest mass probed by average field T dwarfs.

The posterior distribution for the cutoff model peaks more narrowly than the posterior distribution of the two-segment power law, and it too is not strongly dependent on the prior distribution.  Figure \ref{fig:cop} displays the same four altered prior distributions and their resultant posterior distributions as in \S3.3.1, but for the cutoff power law mass function.  The posterior distributions show similar behavior to those of the two-segment power law mass function.  The Bayesian result of this mass function formulation is similar to the previous one, that the data weakly constrain the model parameter values.

\subsubsection{Log normal Mass Functions}

We apply our Bayesian analysis to a third set of model luminosity functions based on a log normal mass function.  As with the power law analyses, there are two free parameters: the characteristic mass, $m_0$, which is allowed to span the range $-1.4 < \log(m_0) < -0.4$; and the width, $\sigma$, which spans 0.35 to 1.35. The nominal prior distribution uses the values of those parameters given by \citet{chab03} ($\log(m_0) = -1.1\pm0.1$ and $\sigma = 0.69\pm0.05$).  The log normal mass functions generate similar luminosity functions to those from power law mass function models through the T dwarf regime, but they diverge at fainter magnitudes (see ${\S}3.3.4$). As noted above, this stems from the turnover in the log normal mass functions at low masses.

The Bayesian analysis of this mass function model yields similar results to those outlined in the previous sections.  Figure \ref{fig:lnp} displays the posterior distributions on $\log(m_0)$ for four variations of the $m_0$ prior distribution.  Unlike the previous analyses, the posterior distribution strongly mirrors the input prior distribution, which means that we cannot constrain the mass function in this case.  The reason for this is that the KCAB data are not at the peak of the log normal distribution.  This effectively means we try to fit the falling slope of the mass function.  The result is that a wide range of possible parameter values are consistent the data.  To properly fit a log normal mass function we need to include data from masses at the peak and above ($>0.10~M_{\odot}$).

\subsubsection{Model Discussion}

All three mathematical representations of the mass function match the data with similar accuracy.  Figure \ref{fig:fit} shows the best fit models from each of the theoretical forms matched against the empirical KCAB densities.  All three luminosity functions are nearly identical to $M_J \sim 17$; only there do they begin to diverge.  The two-segment power law model has slightly higher densities than the log normal model, while the abrupt stop in the cutoff model is the result of the low-mass cutoff.  These differences are all well below the current detection limit, implying that it is very difficult to tell the difference between these models.  This is what our Bayesian output told us.  The posterior distributions on the model parameters are either not well constrained or completely dependant on the prior distribution.

T wo properties of the calibrating KCAB dataset contribute to the weak constraints on the model parameter values. First, the measurements for late-M to mid-L dwarfs \citep{kc03,kc05}, though the most reliable density determinations, fall within trough B of the luminosity function (Figure \ref{fig:fit}).  This region is highly insensitive to changes in the value of the model parameters (Figure \ref{fig:sfmf}).  Second, while the number densities of late-L and T dwarfs depend strongly on the slope of the underlying mass fuction (see \S2.2), their measured space densities have substantial uncertainties.  Consequently, a wide range of parameter values fit the data; and, with the data currently in hand, it is only possible to place weak constraints on the form of the substellar mass function.

\section{The Future: Improved Constraints on the Field Substellar Mass Function}

Since the field substellar mass function is weakly constrained with existing data, further observational efforts must be undertaken.  There are both current and future projects that can improve the mass function constraints.  Follow-up observations of either SDSS or 2MASS sources account for nearly all of the currently known T dwarfs, and that work is continuing (e.g., \citet{burg04}).  With a substantial fraction of both surveys analyzed and their lower apparant magnitude detection limits fully probed, neither survey will extend coverage to significantly lower luminosities.  However, they will continue to bolster the statistics of late-L and T dwarfs.  The Spitzer Space Telescope is capable of carrying out wide-angle surveys for T dwarfs and discovering objects even cooler.  The predicted mid-infrared colors of L, T, and cooler objects are distinctive \citep{bur03,mar}. Though Spitzer is not a survey instrument, several current programs provide large sky coverage.  In particular, two Legacy programs, the Spitzer Wide-area Infrared Extragalactic Survey (SWIRE) and the Galactic Legacy Infrared Mid-Plane Survey Extraordinaire (GLIMPSE), survey 70 and 240 square degrees respectively \citep{swire,glimpse}.

We use the modeling techniques outlined in \S2 to predict the likely number of T and cooler dwarfs detectable by these two Legacy surveys. The Infrared Array Camera (IRAC) on Spitzer provides data in four passbands at 3.6, 4.5, 5.4, and 8.0 $\mu$m \citep{irac}.  The luminosity functions we generate in \S2.3 for $M$-band (4.8 $\mu$m) are roughly equivalent to IRAC Channel-2.  For the purposes of generating a rough estimate of ultracool dwarf detections rates in the SWIRE and GLIMPSE Legacy surveys, the differences between $M$-band and IRAC Channel-2 are not significant enough to compromise our results.  The expected magnitude limits ($5\sigma$) in IRAC Channel-2 for SWIRE and GLIMPSE are 18.8 (5.3 $\mu$Jy) and 15 (185 $\mu$Jy) respectively.  

We calculate the volume limits at each absolute magnitude to estimate the number of T dwarf and cooler objects likely to be detected by each survey.  The apparent magnitude limits of the two surveys set the effective distance limit and, hence, the volume searched.  We limit analysis to T (approximate limits of $11.5 < M_M < 13.5$) and cooler (M$_M > 13.5$) dwarfs, and examine two possible underlying mass functions; an optimistic two-segment power law with $\alpha_2 = 0.8$ and a pessimistic one with $\alpha_2 = 0.0$, both for $m_{12} = 0.09~M_{\odot}$.  With the optimistic model, SWIRE finds $\sim$1100 objects, and $\sim$800 with the pessimistic.  This difference is large enough that SWIRE may be able to perform some rudimentary mass function studies just from the number counts of potential very cool brown dwarfs.

The left side of Figure \ref{fig:swire} displays the expected number of detections (top panel) and the distance limits (bottom panel) for the SWIRE survey, as a function of M$_M$.  Most of the objects ($\sim$1000 of 1100) found will be be T dwarfs.  Moreover, the overwhelming majority of both T and cooler dwarfs will have distances exceeding 50 pc, rendering follow-up observations difficult with currently available instruments.  For example, a bright dwarf cooler than T with M$_M = 13.5$ and $M_J$ = 17 has an apparent magnitude of $m_J$ = 20.5 at 50~pc.  Even with the NIRSPEC instrument on the Keck telescope, it takes several hours to obtain a near-infrared spectrum with signal to noise ratio of 5 \citep{mcl}.

The right side of Figure \ref{fig:swire} displays a similar analysis for GLIMPSE coverage and sensitivity.  We predict a total of $\sim$ 20 objects for the optimistic model and $\sim$13 for the pessimistic, most of which will be T dwarfs with one or two cooler dwarfs, significantly fewer than in SWIRE.  The GLIMPSE estimates are much lower because of the shallower sensitivity, and the larger areal coverage does not make up for that loss.

An all-sky, moderately deep, mid-infrared survey must be undertaken to find large numbers of easily-recoverable, cool brown dwarfs.  This type of survey offers the prospect of finding hundreds of nearby, extremely cool brown dwarfs.  The Wide-field Infrared Survey Explorer (WISE) \citep{wright} is one such mission that has already been proposed, surveying the entire sky at 3.5, 4.7, 12 and 23 $\mu$m.  WISE makes use of the extremely red 3.5$\mu$m--4.7$\mu$m color of very cool dwarfs to distinguish them from other sources.  The predicted 4.7$\mu$m sensitivity will be similar to the SWIRE Spitzer Legacy survey, but will cover the entire sky, not just 70 square degrees.  Overall, WISE will detect over 270,000 T dwarfs and 40,000 cooler objects, including hundreds within 20-30 parsecs of the Sun.

In the near future upcoming ground based facilities will provide the capability of detecting significant numbers of T and cooler objects.  For example, the VISTA collaboration will have a wide-field near-infrared imager behind a 4m class telescope in Chile within 2 years.  Much of the time (75\%) will be allocated to surveys.  This instrument will probe 5-6 magnitudes deeper than 2MASS or DENIS, which enables the compilation of a larger sample of late-T dwarfs and the possible detection of cooler objects.

In the interim, the conjunction of the completed 2MASS survey and the ongoing SDSS provide the best prospects for further understanding of the substellar mass function.  While SDSS does not cover the whole sky as 2MASS does, it supplies critical color information that allows easier extraction of late-L and early-T dwarfs.  The 2MASS colors of late-L and early-T dwarfs fall into a very crowded area of the color-magnitude diagram.  Selection criteria for early-T dwarfs become much cleaner with the addition of short wavelength SDSS photometry (Figure \ref{fig:cmd}).  In this way, we obtain preliminary space densities for late L and T dwarfs before a definitive project, like WISE, is carried out.

\section{Conclusions}

This paper produces new models of the luminosity function of field brown dwarfs.  Through these models, we explore the role of the rapid evolution of brown dwarfs on the luminosity and $T_{eff}$ distributions, and find that those distributions are surprisingly insensitive to changes in the underlying age distribution.  Our main goal is to use our Bayesian statistical method to constrain the field substellar mass function using data on late-M, L, and T dwarfs (KCAB).  

We present results that demonstrate the extant data provide modest constraints on parameter values for the substellar mass function and do not discriminate between commonly used functional forms.  The Bayesian constraints on the three forms of the underlying substellar mass function are as follows:
\begin{itemize}
\item{Two-segment power law: Weak constraints on model parameters with the 60\% confidence limits yielding a range of -0.5 to 0.5 for $\alpha_2$ and $0.04~M_{\odot}$ to $0.10~M_{\odot}$ for $m_{12}$}
\item{Low-mass cutoff power law: Equally weak fit as the two-segment power law, but consistent with a single power law mass function from $0.10~M_{\odot}$ with $\alpha_2 = 0.3\pm0.6$}
\item{Log normal mass function: Existing data do not provide any further constraint on the characteristic mass ($\log{m_0}$)}
\end{itemize}
Through the use of these three model mass functions and our Bayesian analysis we demonstrate that the field substellar mass function cannot be well constrained with existing data.  This is for two reasons: 1) The best quality data, the L dwarf space densities, fall in a trough of the luminosity function that is insensitive to changes in underlying mass function and age distribution models, rendering model constraints using L dwarfs weak at best.  2) The more parameter-sensitive late-L and T dwarfs do not yet have well defined space densities, due to small number statistics and the lack of a volume complete sample.  Therefore, we conclude that improved constraints on the field substellar mass function require further data on cool brown dwarfs.

We predict the cool brown dwarf sensitivities in two Spitzer Legacy surveys with the space densities we obtain from our model luminosity functions.  SWIRE will detect hundreds of T dwarfs and dozens of cooler dwarfs, while GLIMPSE will see only $\sim$20 T dwarfs and at most one cooler object.  However, most of these discoveries will lie at distances greater than 50~pc and thus be too faint for existing instruments to recover efficiently.  We conclude that the best constraints on the field substellar mass function will require large sky surveys at mid-infrared wavelengths that will find hundreds of nearby T and cooler dwarfs.

Thanks to Kelle Cruz and Adam Burgasser for making their data available for use in this paper.  The authors thank Adam Burrows for the use of the Arizona group's models and answering questions regarding them.  P.R.A. also thanks Kelle Cruz for help in proofreading this manuscript and discussing the concepts within.  P.R.A. acknowledges support by a grant made under the auspices of  the NASA/NSF NStars initiative, administered by JPL, Pasadena, CA.  D.E.T.\ acknowledges support from a Space Telescope Science Institute grant to G.\ Bernstein and by NASA through contract number 960785 (Spitzer/MIPS) issued by JPL/Caltech.

\clearpage

\begin{deluxetable}{cccc}
\tablewidth{0pt}
\tablecaption{Power Law Mass Function Estimates}
\tablehead{
\colhead{Location} & \colhead{$\alpha$} & \colhead{Age (Myr)} & \colhead{Reference}
}

\startdata

$\sigma$ Ori               & $0.8  \pm 0.4$  & $\sim5$   & \citep{bej}         \\
$\alpha$ Per               & $0.59 \pm 0.05$ & $\sim90$  & \citep{bn02}        \\
Pleiades                   & $0.60 \pm 0.11$ & $\sim120$ & \citep{mor}         \\
M35                        & $0.18 \pm 0.12$ & $\sim160$ & \citep{bn01b}       \\
Taurus                     & $\sim0.4$       & $\sim1$   & \citep{bri02,lu04}  \\
IC 348                     & $\sim0$         & $\sim2$   & \citep{lu03}        \\
Orion Nebula Cluster       & $\sim0$         & $\sim3$   & \citep{mlla}        \\
`Average Field'            & $0.3  \pm 0.7$  & \nodata   & \citep{kru02}       \\
Field M and L Dwarfs       & $1.5  \pm 0.5$  & \nodata   & \citep{inr99}       \\
Ultracool Field M  Dwarfs  & $<2$            & \nodata   & \citep{inr022}      \\
Field T Dwarfs             & $0.75 \pm 0.25$ & \nodata   & \citep{burgp}       \\

\enddata

\end{deluxetable}

\clearpage

\begin{deluxetable}{ccc}
\tablewidth{0pt}
\tablecaption{Spectral Type Physical Parameters}
\tablehead{
\colhead{SpT} & \colhead{Mean Age (Gyr)} & \colhead{Mean Mass ($M_{\odot}$)}
}

\startdata

M6      & 5.3 & 0.093 \\
L0      & 4.1 & 0.074 \\
L5      & 3.3 & 0.067 \\
Late-L  & 2.9 & 0.063 \\
Early-T & 3.1 & 0.058 \\
Late-T  & 4.8 & 0.048 \\

\enddata

\end{deluxetable}

\clearpage

\newpage
\begin{deluxetable}{ccc}
\tablewidth{0pt}
\tablecaption{Field Space Density Data}
\tablehead{
\colhead{M$_J$} & \colhead{Space Density} & \colhead{Volume Observed}\\
\colhead{} & \colhead{($10^{-3}$pc$^{-3}$)} & \colhead{($pc^{3}$)}
}

\startdata

(10.75)\tablenotemark{a} & 2.50$\pm$0.60 & 13400 \\
11.25   & 1.49$\pm$0.33 & 13400 \\
11.75   & 0.97$\pm$0.27 & 13400 \\
12.25   & 0.75$\pm$0.24 & 13400 \\
12.75   & 0.37$\pm$0.17 & 13400 \\
13.25   & 0.25$\pm$0.13 & 13400 \\
13.75   & 1.00$\pm$0.30 & 13400 \\
(14.25)\tablenotemark{a} & 1.70$\pm$1.42 & 13400 \\
(14.75)\tablenotemark{a} & 2.31$\pm$1.43 & 13400 \\
(15.25)\tablenotemark{a} & 1.90$\pm$0.90 &  3000 \\
15.75   & 2.00$\pm$1.50 &  1250 \\
16.25   & 4.70$\pm$3.00 &   660 \\

\enddata

\tablenotetext{a}{These data not used in the analysis due to incompleteness}

\end{deluxetable}

\clearpage

\begin{figure}
\rotatebox{-90}{
\epsscale{0.8}
\centering
\plotone{f1.ps}}
\figurenum{1}
\label{fig:fad}
\caption{Four empirically derived field age distributions: a) high z star formation \citep{pas}; b) stellar activity in G and K stars \citep{sod}; c) stellar activity in M dwarfs \citep{pmsu3}; d) statistically derived from the color-magnitude diagram of nearby early type stars \citep{hern} (solid lines).  Plotted over these are three idealized age distributions: Uniform (flat dotted line), decreasing (negative slope dotted line), and increasing (positive slope dotted line) star formation rates.  The age is plotted such that 10 is ten billion years ago and 0 is the present.  Each distribution yields the probability of a field star having a particular age.}
\end{figure}

\clearpage

\begin{figure}
\rotatebox{-90}{
\epsscale{0.8}
\centering
\plotone{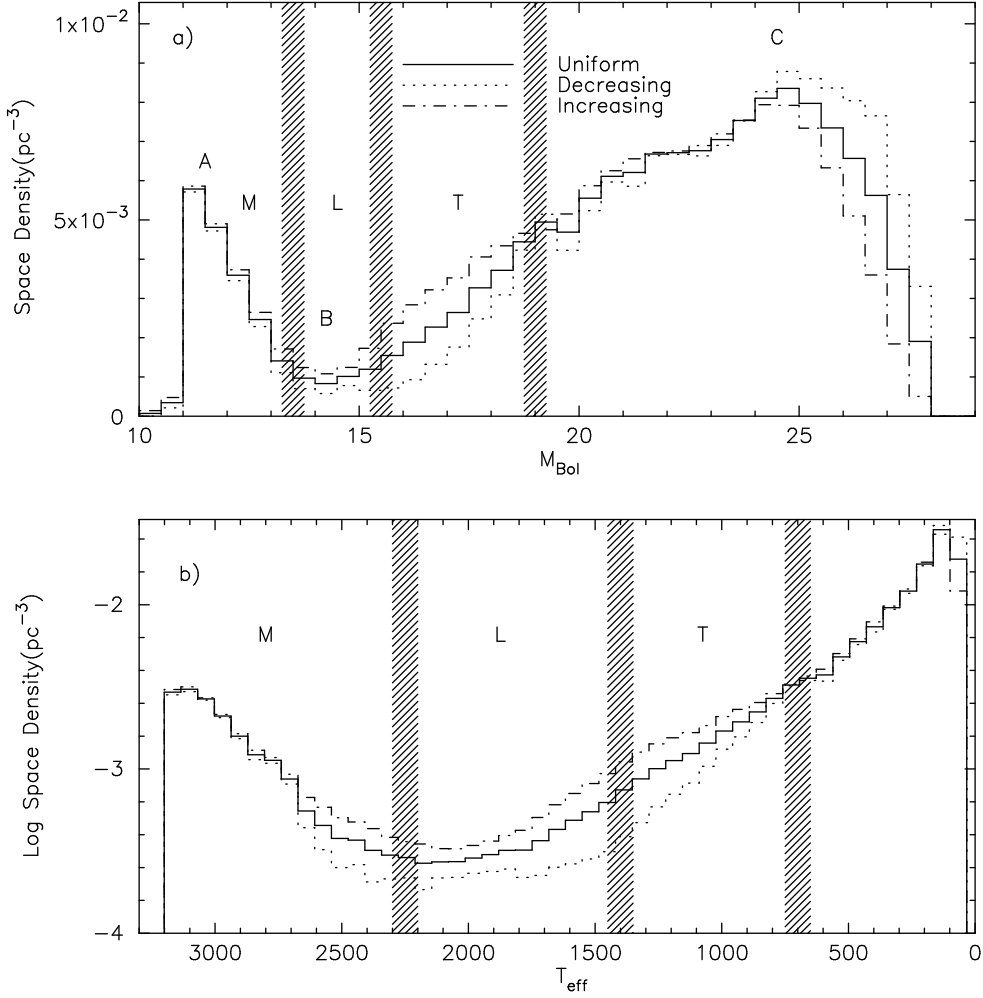}}
\figurenum{2}
\label{fig:asc}
\caption{Model bolometric luminosity functions and $T_{eff}$ distributions derived using three different age distributions; uniform (solid), decreasing (dotted), and increasing (dot-dashed), and a two segment power law with $\alpha_2 = 1$.  Regions in $M_{bol}$ and $T_{eff}$ that correspond to M, L, and T spectral type are delimited by vertical slashed rectangles.  The cutoff in the luminosity function and $T_{eff}$ distribution at bright magnitudes and high temperature is due to the upper mass cutoff of the the \citet{bur} models, and the cutoff at low temperatures and faint magnitudes to the low-mass cutoff of the Burrows models.}
\end{figure}

\clearpage

\begin{figure}
\rotatebox{-90}{
\epsscale{0.8}
\plotone{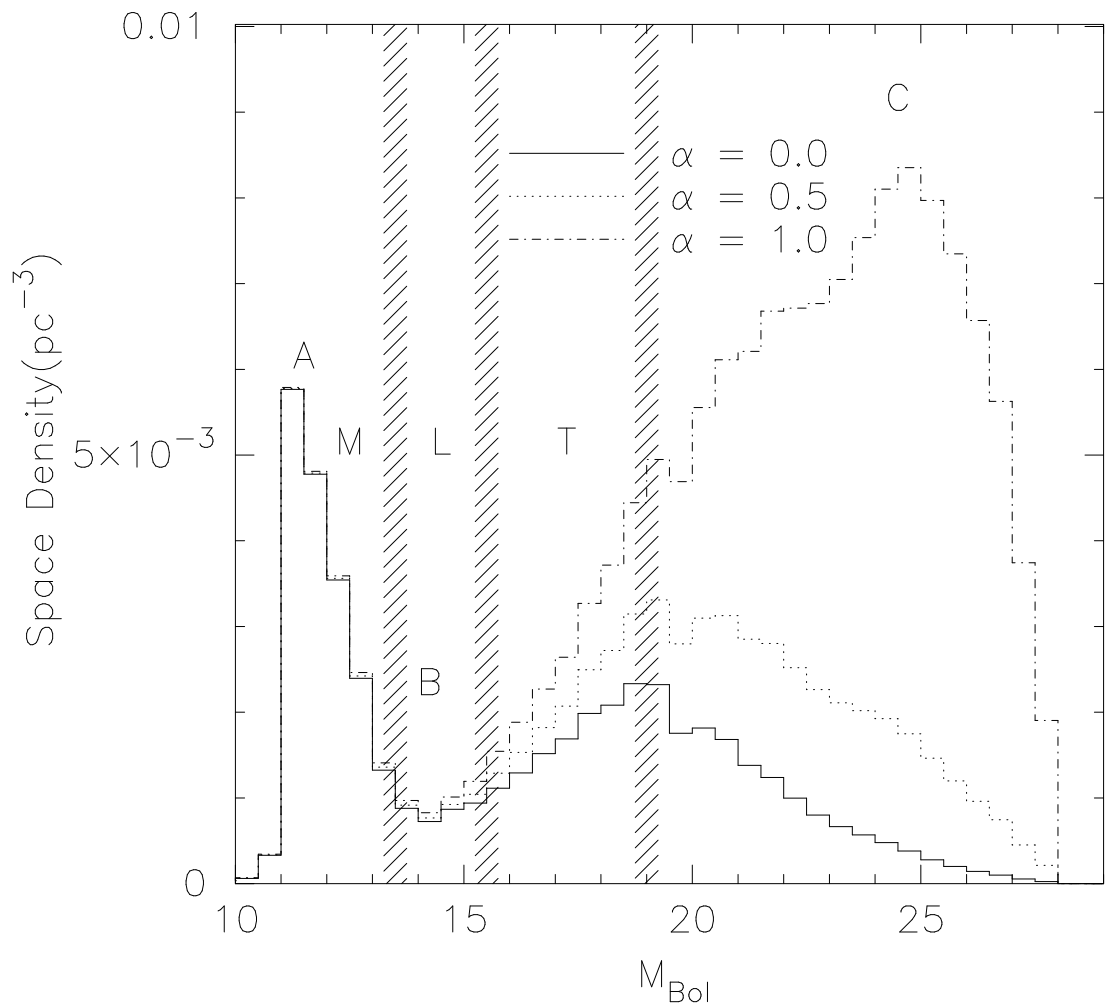}}
\figurenum{3}
\label{fig:sfmf}
\caption{Three bolometric luminosity functions comparing model mass functions for $\alpha_2$ = 0.0, 0.5, and 1.0 and $m_{12} = 0.09~M_{\odot}$, with the same spectral type boundaries as in Figure \ref{fig:asc}a.  Observable variations in the underlying mass function are most apparent in the T dwarf regime.}
\end{figure}

\clearpage

\begin{figure}
\rotatebox{-90}{
\epsscale{0.8}
\plotone{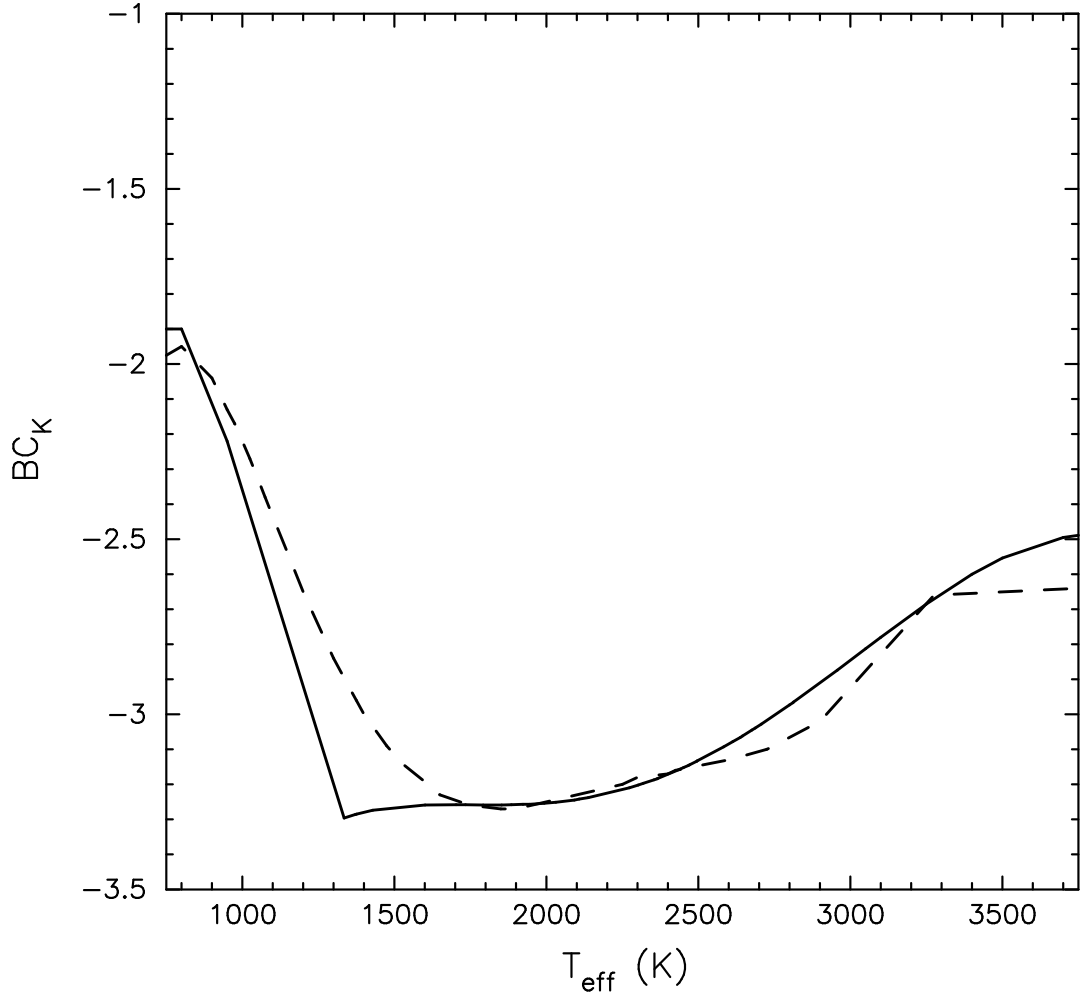}}
\figurenum{4}
\label{fig:bck}
\caption{Comparison of the $K$-band bolometric corrections used in A03 (solid, after \citet{dahn}) and those used here (dashed, from \citet{golim}).  Note the smoother transition around 1300K (the L/T transition) provided by the Golimoski corrections compared to the Dahn corrections.}
\end{figure}

\clearpage

\begin{figure}
\rotatebox{-90}{
\epsscale{0.8}
\plotone{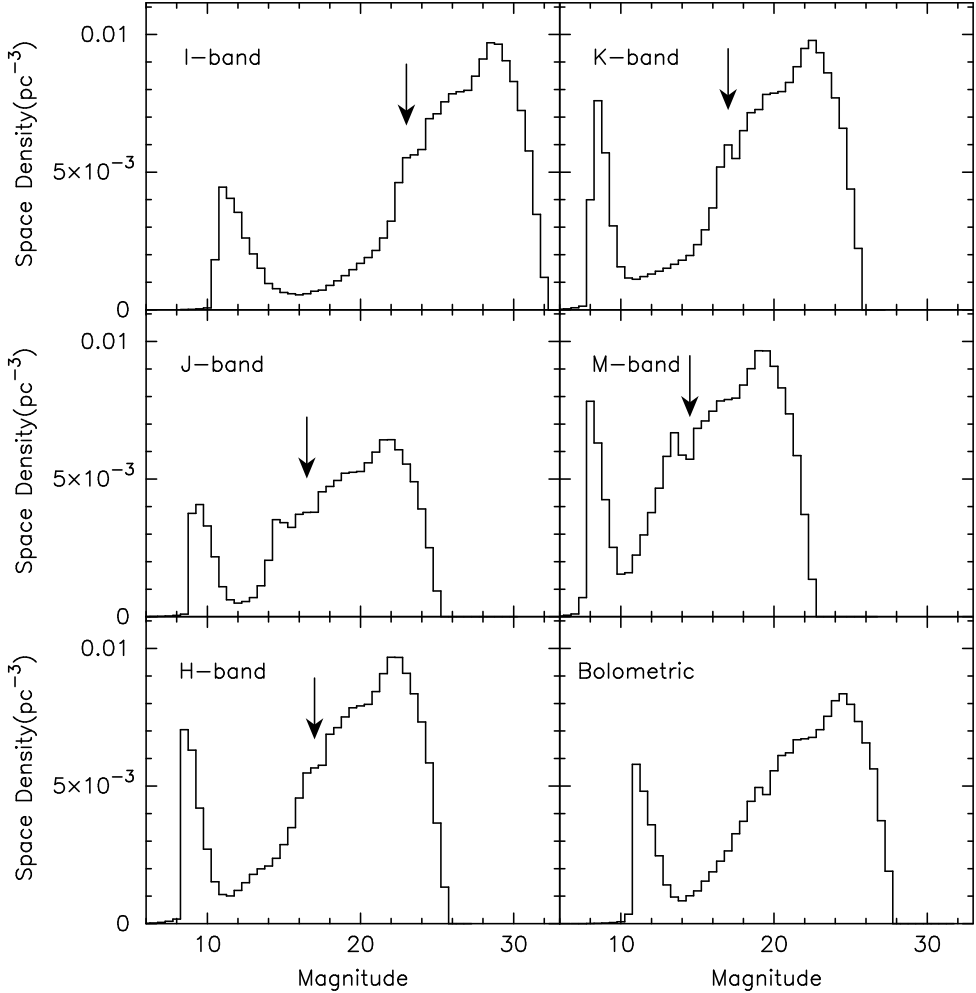}}
\figurenum{5}
\label{fig:bblfs}
\caption{$IJHKM$-band and bolometric magnitude field luminosity functions for a uniform age distribution with $\alpha_2$ = 1.0 and $m_{12}$ = 0.09$M_{\odot}$.  The extra dip or leveling of the broad-band luminosity functions at faint magnitudes (marked by arrows) is caused by the end of the empirical bolometric corrections in the T dwarf regime and the beginning of our extrapolation.  The shape and features in the broad-band luminosity functions at fainter magnitudes than the arrows consequently have larger uncertainties than brighter features.}
\end{figure}

\clearpage

\begin{figure}
\rotatebox{-90}{
\epsscale{0.9}
\plotone{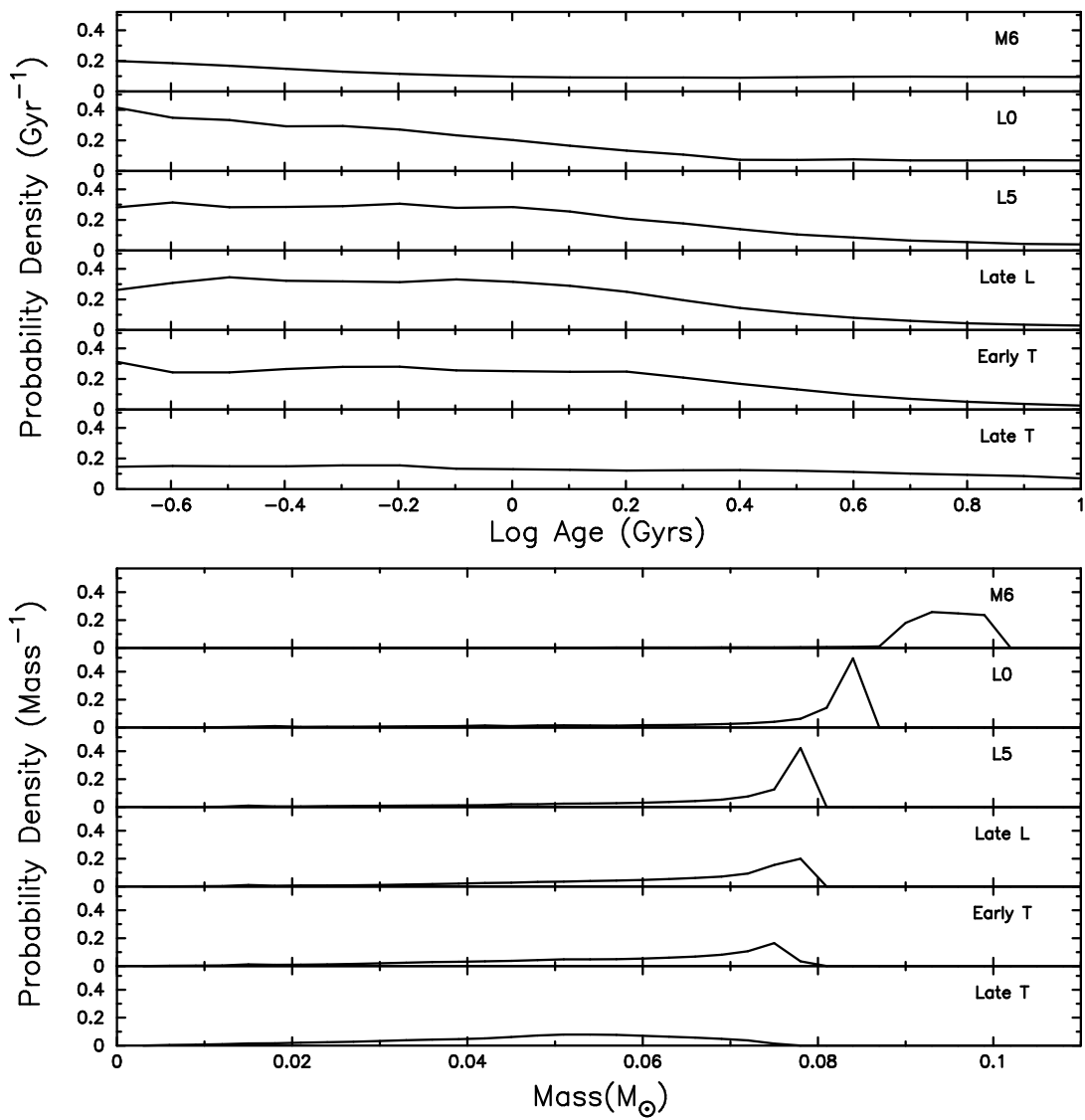}}
\figurenum{6}
\label{fig:spta}
\caption{Predicted age and mass distributions as a function of spectral type for the local field population, derived from the \citet{bur} models and our nominal model mass-age distribution, for M6s, L0s, L5s, late-Ls (L6--L8), early-Ts (T0--T4), and late-Ts (T5--T8).  See Table 2 for the average age and mass of each distribution.}
\end{figure}

\clearpage

\begin{figure}
\rotatebox{-90}{
\epsscale{0.45}
\plotone{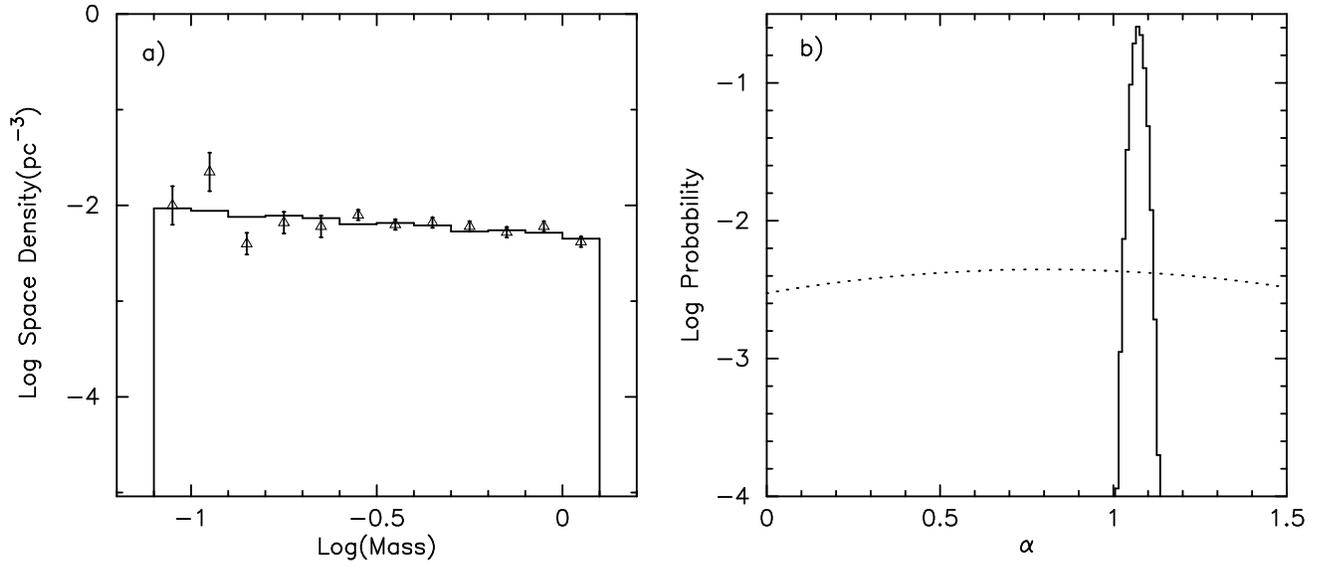}}
\figurenum{7}
\label{fig:pf}
\caption{a) Best fit model mass function (histogram) to the RGH data (triangles) with $\alpha$ = 1.09 $\pm$ 0.017.  b) Bayesian posterior distribution on $\alpha$ for the RGH data fit (solid) and the input prior distribution (dashed).}
\end{figure}

\clearpage

\begin{figure}
\centering
\rotatebox{90}{
\epsscale{0.80}
\plotone{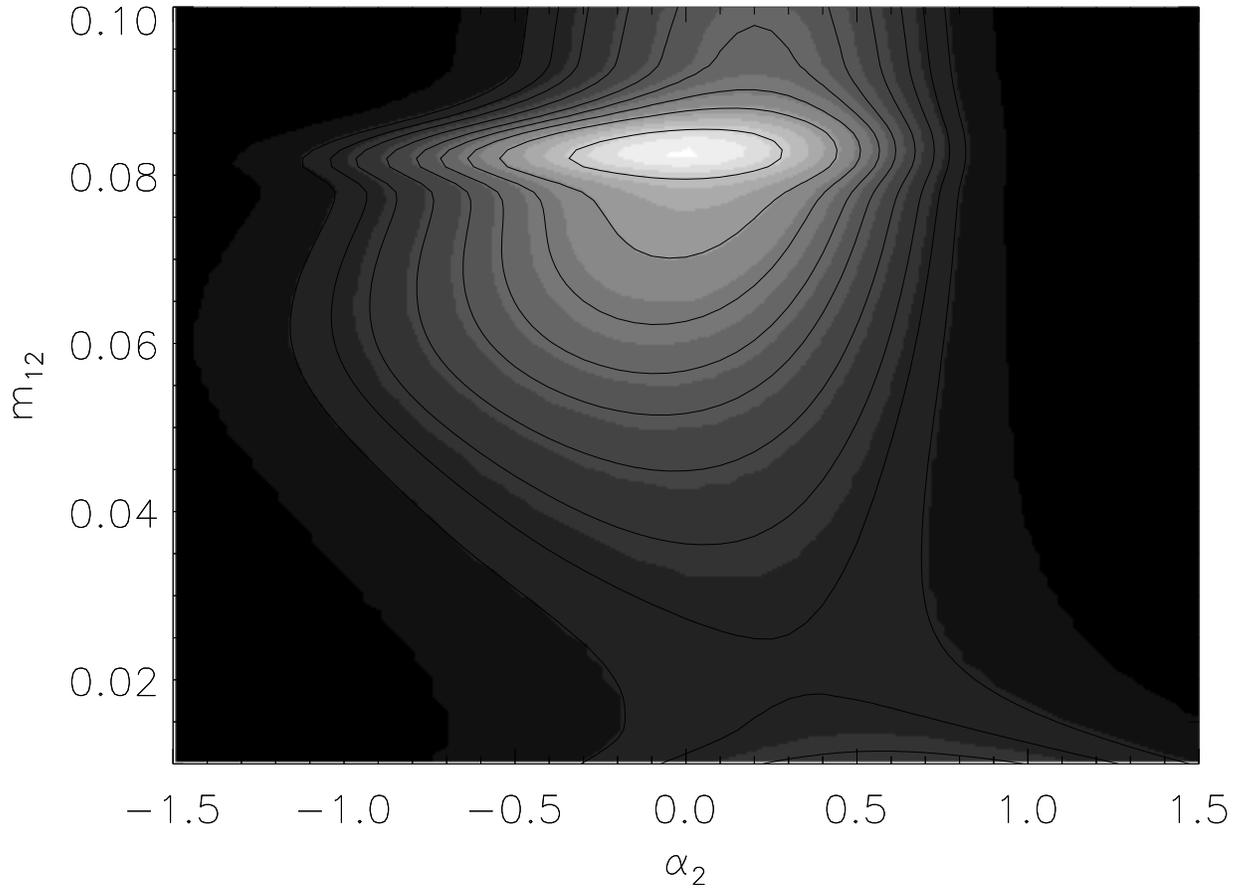}}
\figurenum{8}
\label{fig:ka4}
\caption{Bayesian posterior distribution for the two-segment power law mass function parameters, $\alpha_2$ and $m_{12}$, fits to the KCAB dataset.  White indicates high probability and black low, and the contours are 10\% confidence intervals.}
\end{figure}

\clearpage

\begin{figure}
\centering
\rotatebox{-90}{
\epsscale{0.75}
\plotone{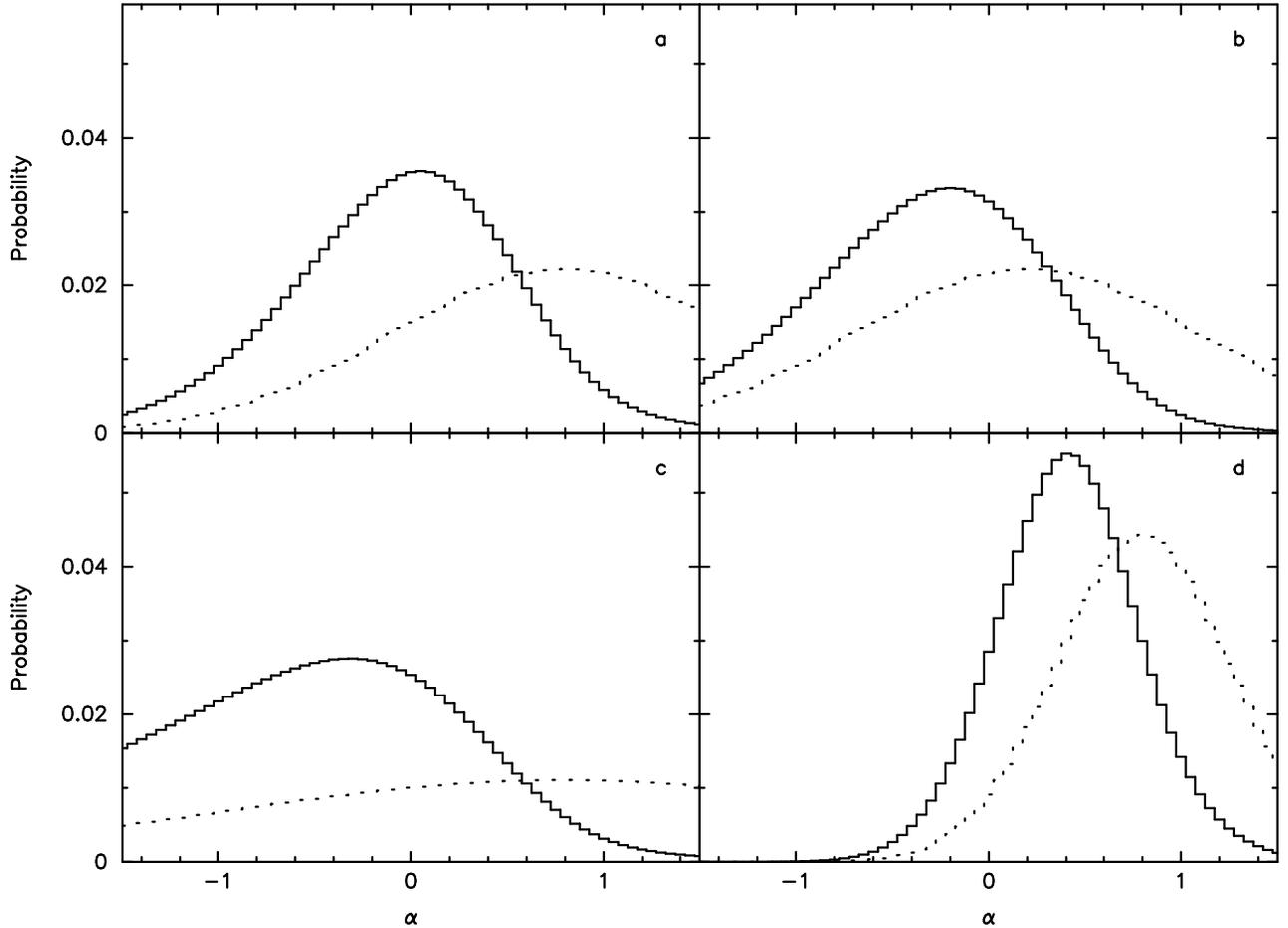}}
\figurenum{9}
\label{fig:twoplp}
\caption{Four Bayesian posterior (solid) and prior (dotted) distributions from the KCAB dataset fit for the two-segment power law mass function parameter $\alpha_2$.  The different prior distributions are as follows: a) Nominal, ${\alpha_2}_0 = 0.8$ and $\sigma_{\alpha} = 0.9$; b) Shifted ${\alpha_2}_0 = 0.2$ and $\sigma_{\alpha} = 0.9$; Widened, ${\alpha_2}_0 = 0.8$ and $\sigma_{\alpha} = 1.8$; d) Narrowed, ${\alpha_2}_0 = 0.8$ and $\sigma_{\alpha} = 0.45$.}
\end{figure}

\clearpage

\begin{figure}
\centering
\rotatebox{90}{
\epsscale{0.80}
\plotone{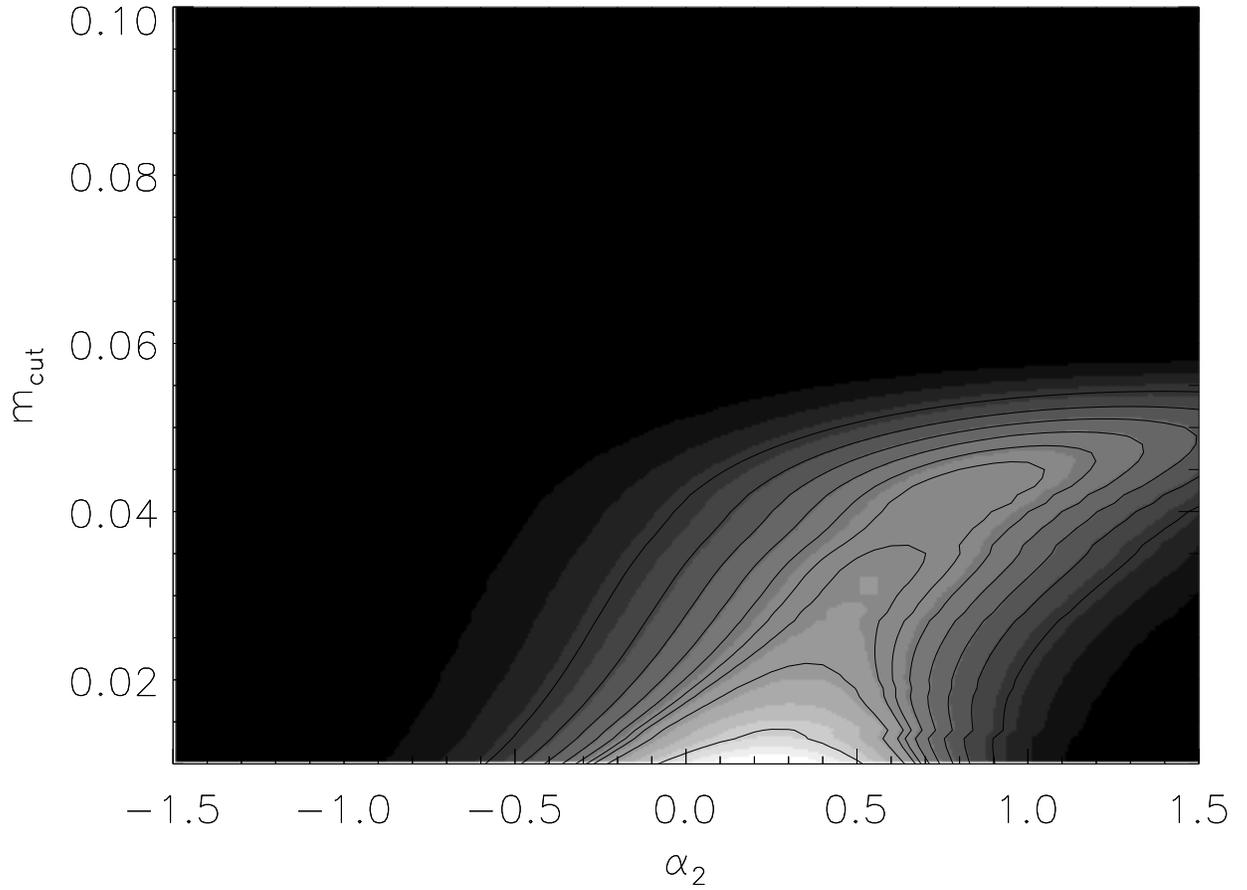}}
\figurenum{10}
\label{fig:ka5}
\caption{Bayesian posterior distribution for the low-mass cutoff power law mass function parameters, $\alpha_2$ and $m_{cut}$, fits to the KCAB dataset.  White indicates high probability and black low, and the contours are 10\% confidence intervals.}
\end{figure}

\clearpage

\begin{figure}
\centering
\rotatebox{-90}{
\epsscale{0.75}
\plotone{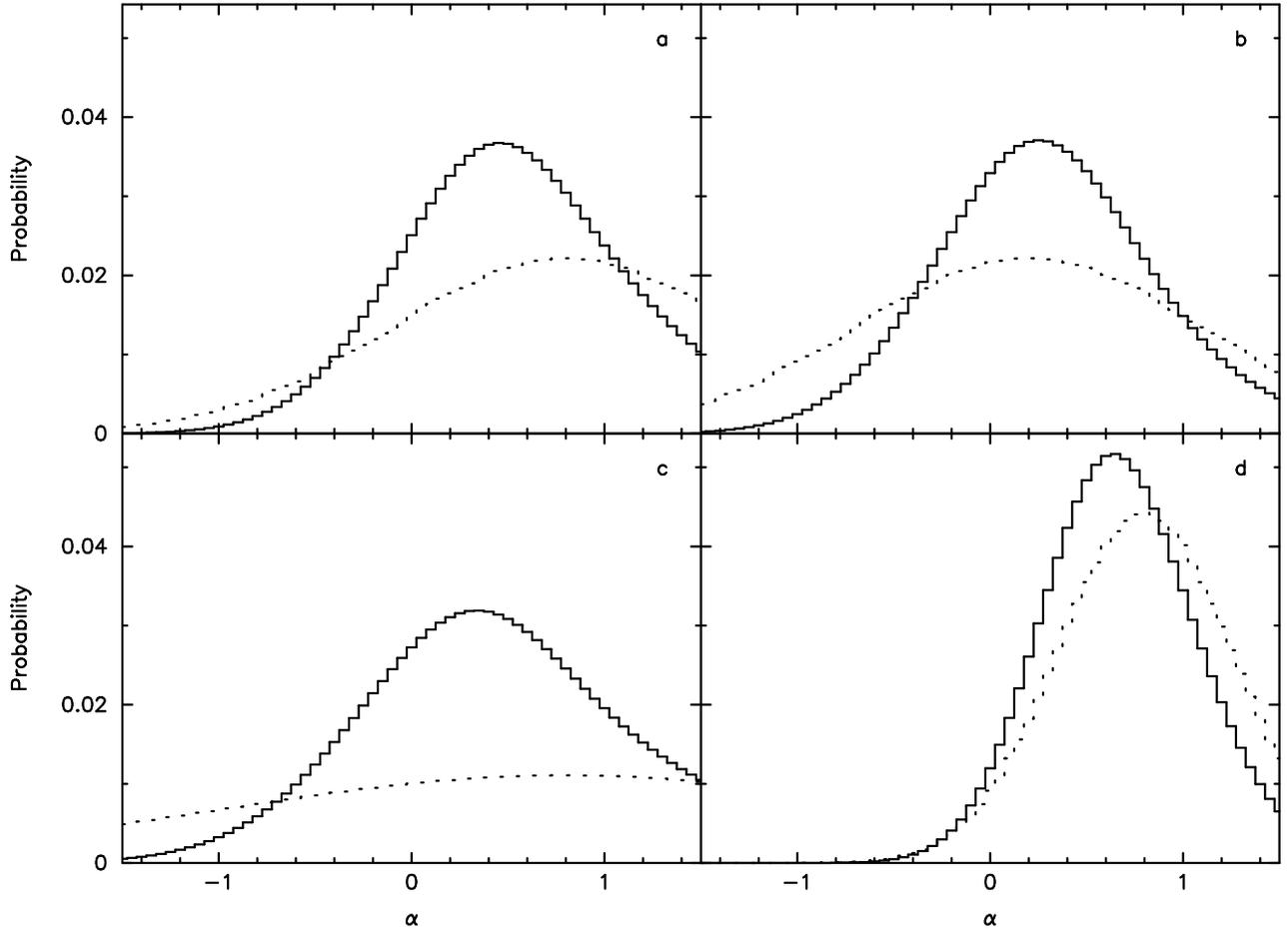}}
\figurenum{11}
\label{fig:cop}
\caption{Four Bayesian posterior (solid) and prior (dotted) distributions from the KCAB dataset fits for the low-mass cutoff power law mass function parameter $\alpha_2$.  The different prior distributions are identical to those used on the two-segment power law (Figure \ref{fig:twoplp}).}
\end{figure}

\clearpage

\begin{figure}
\centering
\rotatebox{-90}{
\epsscale{0.75}
\plotone{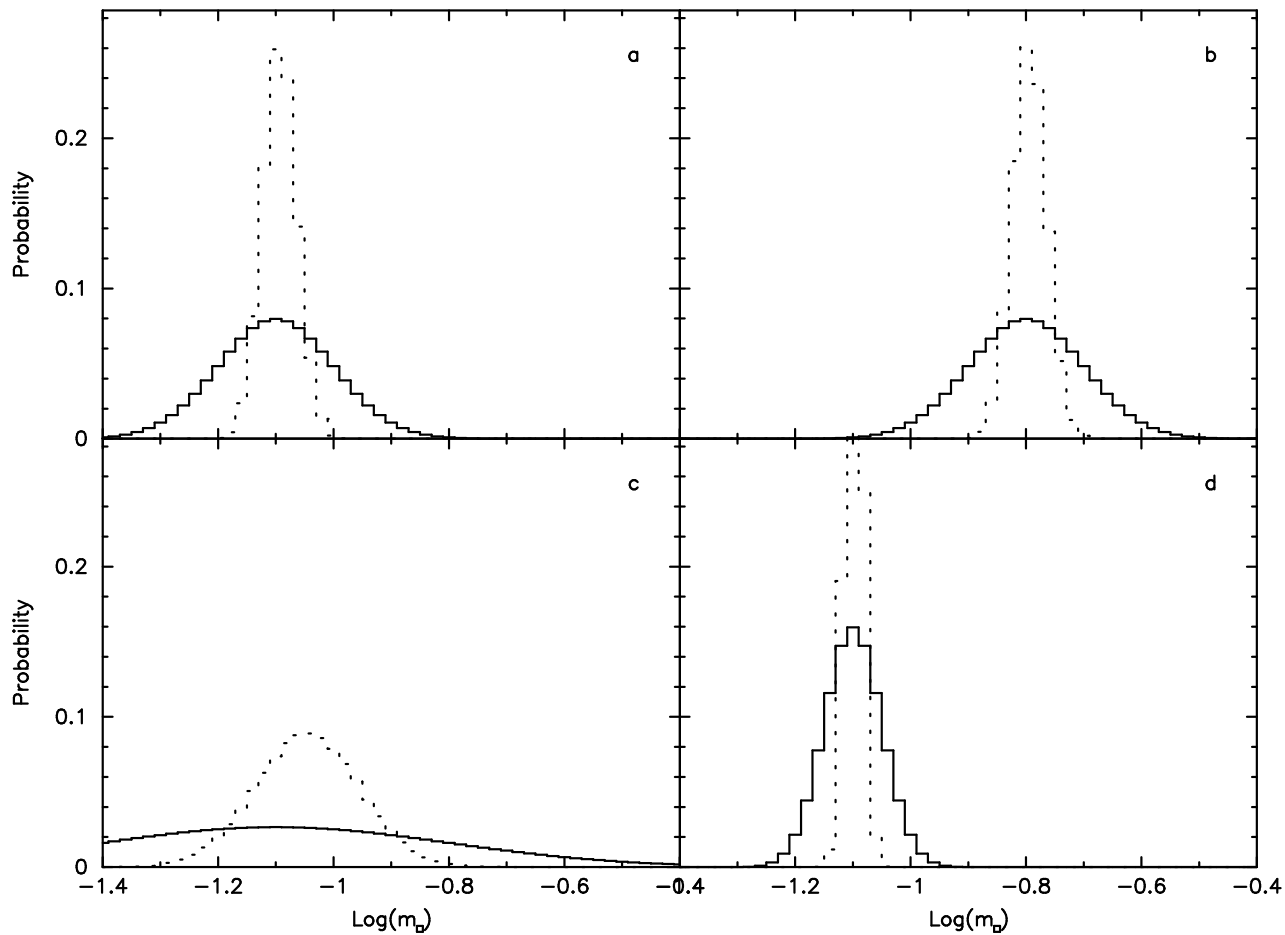}}
\figurenum{12}
\label{fig:lnp}
\caption{Four different posterior (solid) and prior (dotted) distributions for a log normal mass function model parameter $\log(m_0)$ fit to the KCAB dataset.  The different prior distributions are as follows: a) Nominal \citet{chab03} result, $\log(m_0) = -1.1$ and $\sigma_{lm} = 0.1$; b) Shifted, $\log(m_0)$ = -0.8 and $\sigma_{lm} = 0.1$; c) Widened, $\log(m_0) = -0.8$ and $\sigma_{lm} = 0.3$; d) Narrowed, $\log(m_0) = -0.8$ and $\sigma_{lm} = 0.05$.}
\end{figure}

\clearpage

\begin{figure}
\centering
\rotatebox{-90}{
\epsscale{0.75}
\plotone{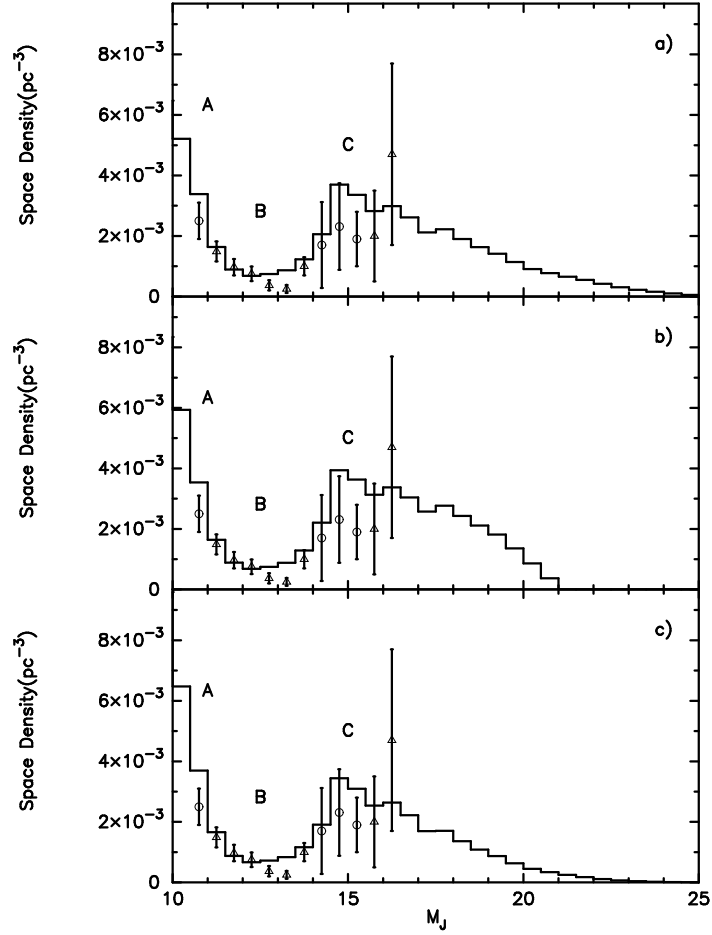}}
\figurenum{13}
\label{fig:fit}
\caption{Joint KCAB observed luminosity function plotted with best fit $J$-band luminosity function models for the three model mass functions: a) two-segment power law mass function with ${\alpha}_2 = 0.0$ and $m_{12} = 0.08~M_{\odot}$; b) low-mass cutoff power law mass function with ${\alpha}_2 = 0.25$ and $m_{cut} = 0.01~M_{\odot}$; c) log normal mass function with $\log(m_0)$ = -1.1 and $\sigma$ = 0.69.  The histograms display the models, the triangular points with error bars are the data used in the fit, and the open circles are data points not used in the fit due to incompleteness (Table 3).}
\end{figure}

\clearpage

\begin{figure}
\centering
\rotatebox{-90}{
\epsscale{0.75}
\plotone{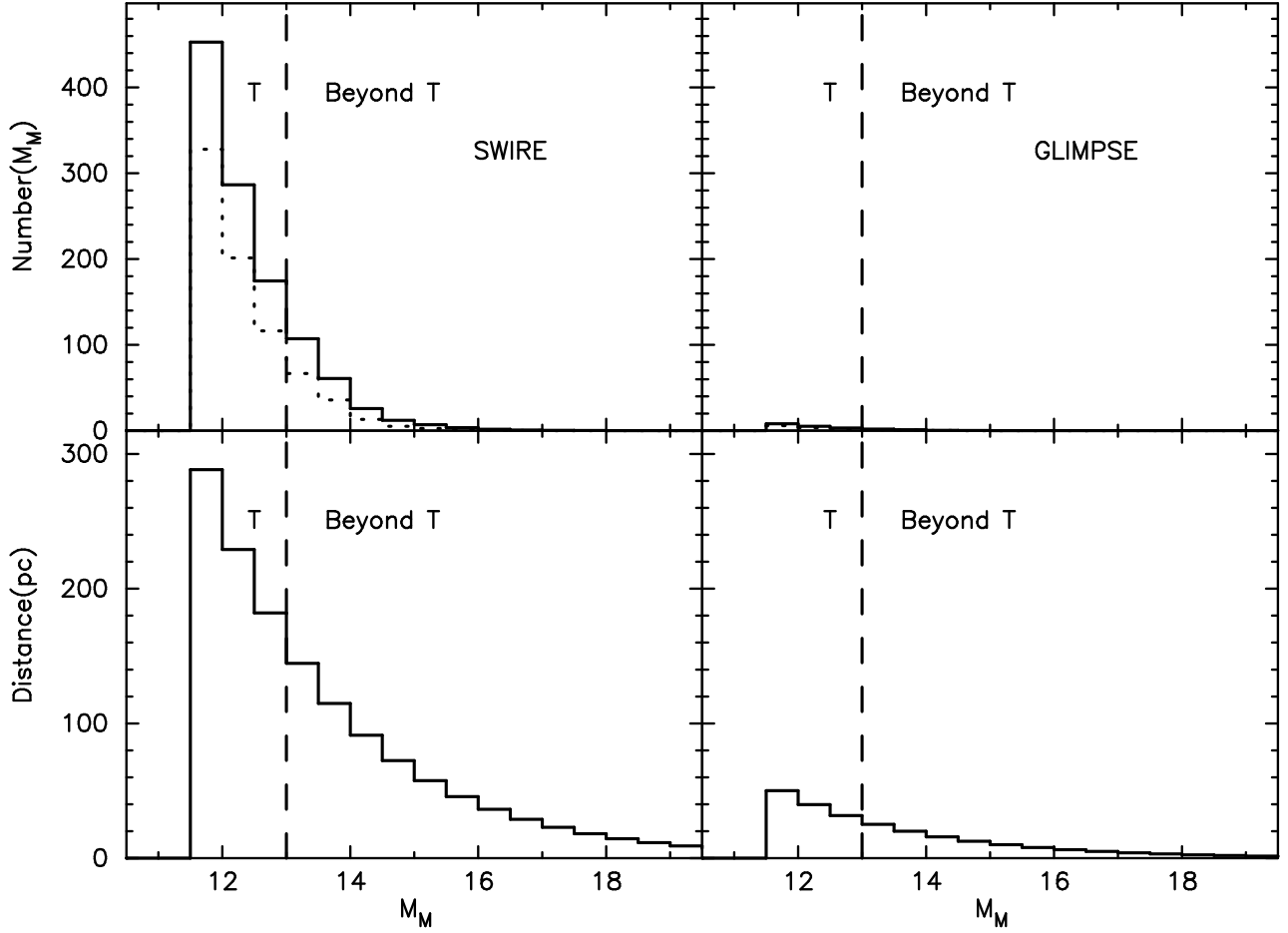}}
\figurenum{14}
\label{fig:swire}
\caption{Expected brown dwarf coverage of the SWIRE (left hand side) and GLIMPSE (right hand side) Spitzer Legacy surveys for two underlying model mass functions.  The solid histogram is for an optomistic two-segment power law mass function with $\alpha_2 = 0.8$, and the dotted histogram is derived from a much shallower mass function with $\alpha_2 = 0.0$.  Top panels display the number of objects predicted given IRAC Channel-2 $5\sigma$ sensitivity limits of 18.8 mag for SWIRE and 15 mag for GLIMPSE.  Bottom panels display the distance coverage of each magnitude bin.  The dashed line at $M_M \sim$ 13.5 marks the expected transition from spectral type T to as-yet unobserved cooler dwarfs.}
\end{figure}

\clearpage

\begin{figure}
\centering
\rotatebox{-90}{
\epsscale{0.5}
\plotone{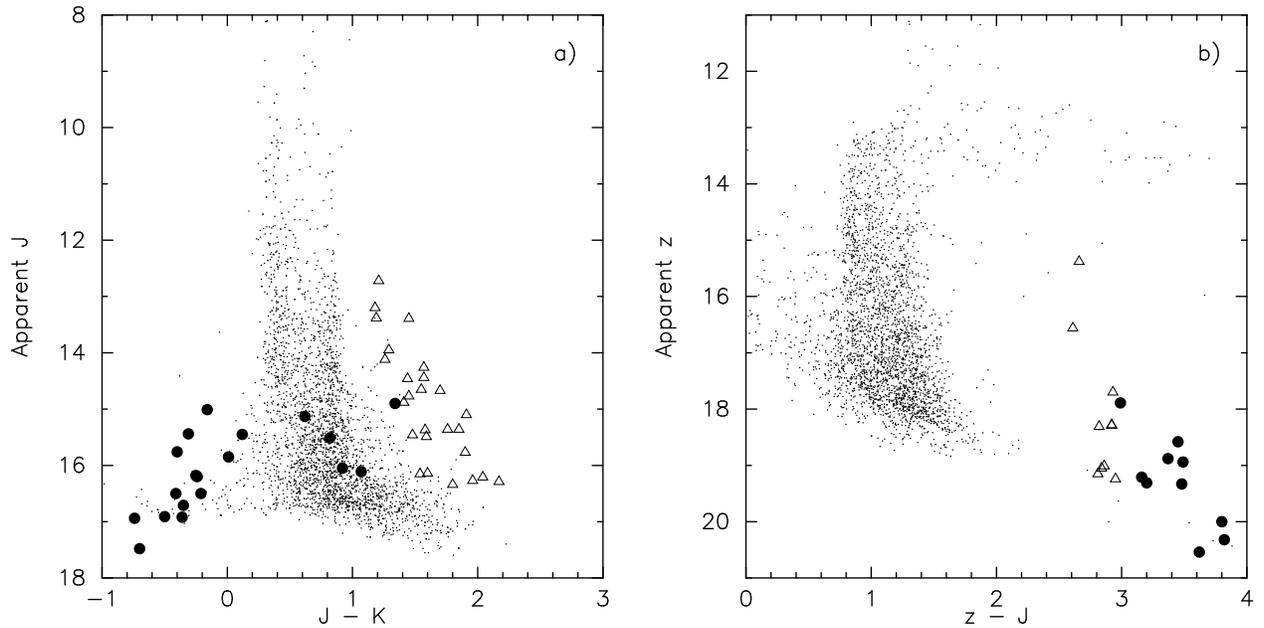}}
\figurenum{15}
\label{fig:cmd}
\caption{a) $J$ vs $J-K$ color-magnitude diagram for a one degree field from 2MASS (small dots) and known L (triangles) and T (solid circles) dwarfs with trigonometric parallaxes (Knapp et~al.\ 2004) shifted to 20~pc (absolute magnitude + 1.51).  Note that the T dwarfs are highly contaminated by background sources.  b) $z$ vs $z-J$ color-magnitude diagram for the same field as a) but using the 2MASS objects that have Sloan $z$-band photometry.  The combination of these surveys provides for cleaner selection of L and T dwarfs.}
\end{figure}

\end{document}